\documentclass[
aps,
prd,        
superscriptaddress,
floatfix,
nofootinbib,
twocolumn,
]{revtex4-1}
\usepackage{ulem}
\usepackage{color}
\usepackage{amsfonts}
\usepackage{amsmath}
\usepackage{amssymb}

\def\be{\begin{eqnarray}}\def\ee{\end{eqnarray}}

\definecolor{brown}{rgb}{0.55,0.27,0.07}

\usepackage{amsmath,amssymb,amsfonts}
\usepackage{graphicx}
\usepackage{bm}         
\usepackage{hyperref}   
\usepackage{physics}    

\newcommand\sect[1]{{\it{#1---}}}

\begin{document}


\title{Hadronic description of nuclear matter and neutron star properties}

\author{Yao Ma}
\email{mayao@nju.edu.cn}
\affiliation{School of Frontier Sciences, Nanjing University, Suzhou 215163, China}

\author{Yong-Liang Ma}
\email{ylma@nju.edu.cn}
\affiliation{School of Frontier Sciences, Nanjing University, Suzhou 215163, China}

\author{Jia-Ying Xiong}
\email{xiongjiaying21@mails.ucas.ac.cn}
\affiliation{School of Fundamental Physics and Mathematical Sciences, Hangzhou Institute for Advanced Study, UCAS, Hangzhou 310024, China}
\affiliation{School of Frontier Sciences, Nanjing University, Suzhou 215163, China}

\date{\today}

\begin{abstract}

The composition of the neutron star is one of the most fundamental and long-standing problems in nuclear- and astro-physics. The known properties of nuclear matter, together with the astronomical observations, impose the stringent and interconnected constraints on the theoretical descriptions. In this work, by using the most general quantum hadrodynamics model including $\sigma, \omega, \rho$ and $a_0$ in addition to nucleons, and performing a Bayesian joint analysis of experimental nuclear matter data, the heavy-ion flow pressure, and astrophysical observations including the mass-radius inferences of PSR J0740+6620, PSR J0437$-$4715, PSR J0614$-$3329 and HESS J1731$-$347 and the tidal posterior of GW170817, we point out that the nuclear matter made of only hadrons can provide a unified description of nuclear matter properties and astrophysical observations.
In addition, we find that the speed of sound in the GQHD develops a non-monotonic, peak-like structure which is absent in the Walecka-type models TM1, NL3 and FSU-$\delta6.7$: in the parameter sets accommodating the small radii of PSR J0614$-$3329 or HESS J1731$-$347 it stays at $v_s^2\sim10^{-2}$ up to $\sim1.5n_{\rm ref}$, with $n_{\rm ref}=0.16$~fm$^{-3}$, rises abruptly between $\sim1.7n_{\rm ref}$ and $\sim2.5n_{\rm ref}$ to $v_s^2\simeq(0.5-0.6)$, well above the conformal value, and then passes through a maximum or levels off. This soft-to-stiff transition, accompanied by a pronounced softening of the symmetry energy, results in small size intermediate mass neutron stars, $R_{1.4}\simeq (11.1-11.4)$~km, together with the maximum mass $(2.2-2.3)M_\odot$, and, to our knowledge, has not been found before in the Walecka-type relativistic mean-field models. What we find here indicate that the sequential measurement of neutron star mass and radius by the next generation facilities, especially that of the intermediate mass neutron stars, is crucial for distinguishing the pure nucleonic stars from the hybrid ones.

\end{abstract}

\maketitle

\allowdisplaybreaks


\section{Introduction}

The properties of dense nuclear matter (NM) especially that relevant to the cores of massive neutron stars (NSs) have been studied for several decades but there are still several fundamental questions, for example, the composition of the matter, are waiting for clarification. The observation of massive NSs~\cite{Demorest:2010bx,Antoniadis:2013pzd} and the detection of gravitational wave (GW) stemming from the binary NS merger~\cite{LIGOScientific:2017vwq} further motivated the interests in this direction~\cite{Baym:2017whm,Annala:2019puf,Ma:2019ery}. The clarification of these questions is crucial for understanding the nuclear force, the deconfinement phase transition, the evolution of compact stars as well as the GWs for the ground based facilities.

The observations of massive NSs around $2M_\odot$~\cite{Demorest:2010bx,Antoniadis:2013pzd} with $M_\odot$ being the solar mass indicate that the equation of state (EoS) of compact star matter should be stiff and sound velocity (SV) favors a peak structure at medium density~\cite{Altiparmak:2022bke}. However, the measurement of PSR J0030+0451~\cite{Vinciguerra:2023qxq} and PSR J0437-4715~\cite{Choudhury:2024xbk}, and especially PSR J0614-3329~\cite{Mauviard:2025dmd}, have collectively shrink the radii of the intermediate mass NS and substantially soften the EoS at intermediate densities. The central compact object in the supernova remnant HESS J1731$-$347, if interpreted as a NS with a carbon atmosphere, is even lighter and smaller, $M\simeq0.77M_\odot$ and $R\simeq10.4$~km~\cite{Doroshenko:2022nwp}, although this inference carries sizable model dependence from the atmosphere and the distance.
This tension makes it increasingly challenging to satisfy all
observational constraints simultaneously within theoretical
frameworks. Moreover,the nuclear physics experiments, such as the heavy-ion collisions (HICs) and the analysis of nuclei structures~\cite{Youngblood:1999zza,Brown:2000pd,Karataglidis:2001yn,Steiner:2005rd,Danielewicz:2002pu,Klahn:2006ir,Li:2008gp,Russotto:2016ucm,Dutra:2014qga,Chen:2010qx,Zhang:2022bni,Huang:2025uzc}, have already provided valuable constraints on the properties of NM around saturation density \(n_0\approx 0.16\ \rm fm^{-3}\), and the flow data of $\rm Au+Au$ collisions constrain the pressure of symmetric NM up to $\sim (4-5)n_0$~\cite{Danielewicz:2002pu}.

In the microscopic description of NM, it is natural to take hadrons---baryons and mesons---as the explicit degrees of freedom (DoFs) at low density. However, at high densities, i.e., $\gtrsim 2 n_0$, the situation is quite unclear since it is naively argued that the overlaps between nucleons cannot be overlooked. In literature, for various purposes and intuitions, the compact star matter are preconceived as quark matter~\cite{Baym:2017whm}, quarkyonic matter~\cite{McLerran:2007qj,McLerran:2018hbz}, color-flavor-locked matter~\cite{Alford:2017qgh,Geissel:2025vnp}, topology objects~\cite{Paeng:2017qvp,Ma:2018jze}, as well as the pure hadronic matter~\cite{Serot:1984ey,Serot:1997xg} to be discussed in this work. Even in the hadronic matter description, a plenty of modifications have been made for various purpose, for example, including the $a_0(980)$ ($\delta$ resonance), $\rho$-$\omega$ mixing, $\sigma$-$\rho$ mixing, etc, for specific purposes. Therefore, it is crucial to clarify whether the simplest approach, here the hadronic description, can saturate all the constraints from nuclear experiments and astro-observations before going to models with exotic DoFs.

To complete our motivation, we consider the most general quantum hadrodynamics (GQHD) model including $\sigma, \rho,\omega$ and $a_0$ in addition to nucleons, written up to dimension-4. The parameters in the model are estimated by using the Bayesian joint analysis (BJA) with respect to the NM properties, the HIC flow pressure and astrophysical observations. Since the M-R inferences of the most compact objects, PSR J0614-3329 and HESS J1731$-$347, are the most demanding and also the most model dependent ones, we construct four parameter sets, GQHD1--GQHD4, which differ only in which of these M-R inferences enter the likelihood. We find that the pure hadronic description can yield the NM properties and NS observations including the most stringent constraints PSR J0614-3329 and HESS J1731$-$347,
and can generate a peak structure of the SV at intermediate densities, without resorting to any transitions from hadron to exotic configurations. Due to the non-monotonic structure of SV which partitions the EoS to soft and stiff segments, the NSs with intermediate mass have smaller radii, consistent with PSR J0614-3329, but maximum mass of NS is $(2.2-2.3)M_\odot$, supporting the massive NS.
Such a peak structure is not produced by the widely used Walecka-type parametrizations, whose SV rises smoothly and almost monotonically; as we show below, the soft-to-stiff structure is present in all four parameter sets of the general model and is most pronounced in those that reproduce the M-R inferences of the compact objects. Which combinations of the dimension-4 operators are responsible for it is a question about the interplay of the full set of couplings rather than of any single term, and is left to a separate systematic study; here we concentrate on the structure itself and on its observable consequences.

This paper is organized as follows. The GQHD model is set up in Sec.~II and the BJA with the data entering it in Sec.~III. Our results---the NM and NS properties, the SV and the tidal deformability---are presented in Sec.~IV, and Sec.~V is devoted to our conclusions. Details of the likelihood and of the parameter sets are given in Apps.~\ref{app:BJA} and~B.

\section{The model}

We consider the most GQHD that incorporates the non-strange scalar and vector mesons below \(1~\mathrm{GeV}\), $\sigma, \omega, \rho$ and $a_0$ mesons, as well as the nucleons. We will not consider pion in the Lagrangian since its mean field vanishes in the parity-even, uniform relativistic mean field (RMF) approximation adopted here. Up to dimension-four, the model takes the form
\be
\mathcal{L} & = & \mathcal{L}_N+\mathcal{L}_\sigma+\mathcal{L}_\omega+\mathcal{L}_\rho+\mathcal{L}_{a_0}+\mathcal{L}_I\ ,
	\label{eq:Lag}
\ee
where
\be
\mathcal{L}_N & = &\bar{\Psi} \left(i\gamma^\mu \partial_\mu - M_N\right) \Psi \ ,\nonumber\\
\mathcal{L}_\sigma & = & \frac{1}{2} \left(\partial_\mu \sigma \partial^\mu \sigma - m_\sigma^2 \sigma^2\right) - \frac{1}{3} g_2 \sigma^3 - \frac{1}{4} g_3 \sigma^4 \nonumber\\
& &{} - \frac{1}{2} g_4 \sigma^2 \omega^\mu \omega_\mu - \frac{1}{2} g_5 \sigma^2 \vec{\rho}^\mu \cdot \vec{\rho}_\mu\nonumber\\
& &{} -g_6\sigma \vec{\rho}^\mu \cdot \vec{\rho}_\mu- g_7 \sigma \omega^\mu \omega_\mu\ ,\nonumber\\
\mathcal{L}_\omega & = &{}-\frac{1}{4} \Omega^{\mu\nu} \Omega_{\mu\nu} + \frac{1}{2} m_\omega^2 \omega^\mu \omega_\mu + \frac{1}{4} c_3 \left(\omega^\mu \omega_\mu\right)^2 \nonumber\\
& &{} + \frac{1}{2} c_4 \omega^\mu \omega_\mu \vec{\rho}^\mu \cdot \vec{\rho}_\mu\ ,\nonumber\\
\mathcal{L}_\rho & = &{} -\frac{1}{4}\vec{P}^{\mu\nu} \cdot \vec{P}_{\mu\nu} + \frac{1}{2} m_\rho^2 \vec{\rho}^\mu \cdot \vec{\rho}_\mu + \frac{1}{4} d_3 \left(\vec{\rho}^\mu \cdot \vec{\rho}_\mu\right)^2\ ,\nonumber\\
\mathcal{L}_{a_0} & = & \frac{1}{2}\partial_\mu \vec{a}_0\cdot\partial^\mu\vec{a}_0-\frac{1}{2}m_{a_0}^2 \vec{a}_0\cdot \vec{a}_0+\frac{1}{4}b_3\left(\vec{a}_0\cdot \vec{a}_0\right)^2\nonumber\\
& &{} +\frac{1}{2}b_4\left(\vec{a}_0\cdot \vec{a}_0\right)\left(\vec{\rho} ^\mu\cdot\vec{\rho}_\mu\right)+\frac{1}{2}b_5\sigma\left(\vec{a}_0\cdot \vec{a}_0\right) \nonumber\\
& &{} +\frac{1}{2}b_6\vec{a}_0\cdot\vec{a}_0\sigma^2+\frac{1}{2}b_7\vec{a}_0\cdot\vec{a}_0\omega^\mu\omega_\mu+b_8\vec{a}_0\cdot\vec{\rho} ^\mu\omega_\mu\sigma\nonumber\\
& &{} +b_9\vec{a}_0\cdot\vec{\rho}_{\mu}\omega^{\mu}\ ,\nonumber\\
\mathcal{L}_I & = & -\bar{\Psi} \left(g_\sigma \sigma + g_\omega \gamma^\mu \omega_\mu + g_\rho \gamma^\mu \vec{\rho}_\mu+g_{a_0}\vec{a}_0\right) \Psi\ .
\ee
In the model, $\Psi = (p,n)^T$ is the iso-doublet of nucleon field,
\(\vec{\rho}_\mu = \rho^i_\mu \tau^i\), \(\vec{a}_0 = a_0 ^i\tau^i\), and \(\Omega^{\mu\nu} = \partial^\mu \omega^\nu - \partial^\nu \omega^\mu\) and \( \vec{P}^{\mu\nu} = \partial^\mu \vec{\rho}^\nu - \partial^\nu \vec{\rho}^\mu \) are field strength tensors of the corresponding vector mesons.

In the standard RMF approximation, the meson fields are replaced by their expectation values, so that the effective proton and neutron masses read
\be
M_p^\ast & = & M_N+g_\sigma\sigma+g_{a_0}a,\nonumber\\
M_n^\ast & = & M_N+g_\sigma\sigma-g_{a_0}a ,
\ee
with $\sigma$ and $a$ being the expectation values of $\sigma$ and $a_0$ meson fields, respectively.
Denoting the proton and neutron scalar densities by \(n_{s,p}\) and \(n_{s,n}\), with
\begin{equation}
\begin{aligned}
n_{s,i}
&=\frac{M_i^*}{\pi^2}\int_0^{k_{F,i}}
\frac{k^2\,dk}{\sqrt{k^2+M_i^{*2}}} \\
&=\frac{M_i^{*3}}{2\pi^2}
\left[
t_i\sqrt{1+t_i^2}-\sinh^{-1}(t_i)
\right], \\
k_{F,i}
&=\left(3\pi^2 n_i\right)^{1/3},
\qquad
t_i=\frac{k_{F,i}}{M_i^*},
\qquad i=p,n ,
\end{aligned}
\end{equation}
the coupled mean-field equations for the \(\sigma\), \(\omega\), \(\rho\), and \(a_0\) fields take the forms
\begin{widetext}
\be
0 & = & \,m_\sigma^2\sigma+g_2\sigma^2+g_3\sigma^3
+g_4\sigma\omega^2+g_5\sigma\rho^2
-b_6a^2\sigma-b_8\omega\rho a
+g_6\rho^2-\frac{1}{2}b_5a^2+g_7\omega^2
+g_\sigma(n_{s,p}+n_{s,n}),\nonumber\\
0 & = & \,m_\omega^2\omega+c_3\omega^3+c_4\omega\rho^2
-g_4\sigma^2\omega+b_7a^2\omega
+b_8\rho\sigma a-2g_7\sigma\omega
+b_9\rho a-g_\omega(n_p+n_n),\nonumber\\
0 & = &\,m_\rho^2\rho+d_3\rho^3+c_4\omega^2\rho
-g_5\sigma^2\rho+b_4a^2\rho
+b_8\omega\sigma a-2g_6\sigma\rho
+b_9\omega a-g_\rho(n_p-n_n),\nonumber\\
0 & = & \,m_{a_0}^2a-b_3a^3-b_4\rho^2a-b_6\sigma^2a
-b_7\omega^2a-b_8\rho\omega\sigma
-b_5a\sigma-b_9\omega\rho
+g_{a_0}(n_{s,p}-n_{s,n}) .
\ee
\end{widetext}

The energy density \(\mathcal{E}\) is obtained within the same mean-field treatment, and the pressure follows from it through the thermodynamic relation $P=-\mathcal{E}+n\,{d\mathcal{E}}/{d n}$, where \(n=n_n+n_p\) is the total baryon number density.
Then the NS mass-radius (M-R) relation and the tidal deformability can be obtained by solving the Tolman-Oppenheimer-Volkoff (TOV) equation and the equation for the quadrupole Love number~\cite{Hinderer:2007mb} with the calculated EOS of beta-equilibrium $npe$ nuclear matter, $\mu_n-\mu_p=\mu_e$ and $n_p=n_e$.
Additionally, the EoS of the NS crust is considered by matching the standard BPS EoS~\cite{Baym:1971pw} to the RMF core through a Maxwell construction, i.e., at equal baryon chemical potential and pressure, with the density discontinuity at the junction kept explicitly in the TOV and Love-number integrations. The same prescription is applied to all EoSs compared below.

Considering the well determined masses of mesons and nucleons~\cite{ParticleDataGroup:2024cfk}, we set \(m_{\omega}=783\ {\rm MeV}\), \(m_{\rho}=770\ {\rm MeV}\), \(m_{a_0}=980\ {\rm MeV}\), and \(M_N=939\ {\rm MeV}\) in the numerical process for the GQHD sets; the reference models keep their published masses (Tab.~\ref{tab:Par}). The mass of \(\sigma\) meson is less certain due to its broad width and strong coupling to pions, so we vary it from \(400\) to \(800\ {\rm MeV}\).
By using the lessons from the chiral nuclear force models~\cite{Ren:2016jna,Djukanovic:2006mc,Girlanda:2010ya,Polinder:2006zh}, we limit the couplings \(g_{\sigma}\), \(g_{\omega}\), \(g_{\rho}\) and \(g_{a_0}\) within \((-20,20)\). From literature~\cite{Ma:2025llw, Kovacs:2021ger, Parganlija:2012fy}, the magnitudes of three-meson couplings and four-meson couplings are taken to be less than \(10000~\rm MeV\) and 1000, respectively.
Then, the 21-dimensional parameter set \(\boldsymbol{\theta}\) to be estimated includes:
\( g_{\sigma}, g_{\omega}, g_{\rho }, g_{a_0}, b_i, c_3, c_4, g_j, d_3\) with $(i=3, \cdots, 9, j=2,\cdots 7)$, and \( m_\sigma\).

\section{The Bayesian analysis}

The Beyesian analysis provides a systematic framework to incorporate various constraints from nuclear physics experiments and astrophysical observations based on Bayes' theorem~\cite{bayes1763essay},
\begin{equation}
\label{eq:Bayes}
    p(\boldsymbol{\theta} \mid \mathbf{D})=\frac{p(\mathbf{D} \mid \boldsymbol{\theta}) p(\boldsymbol{\theta})}{p(\mathbf{D})}\ ,
\end{equation}
where \(\boldsymbol{\theta}\) denotes the model parameters, \(p(\boldsymbol{\theta})\) is the prior distribution encoding theoretical expectations and empirical knowledge about the parameters before considering any data, \(p(\mathbf{D} \mid \boldsymbol{\theta})\) is the likelihood function quantifying the compatibility between the model predictions and the data, and \(p(\mathbf{D})\) is the Bayesian evidence serving as a normalization constant.
\(p(\mathbf{D})\) is defined as
\begin{equation}
    p(\mathbf{D})=\int p(\mathbf{D} \mid \boldsymbol{\theta}) p(\boldsymbol{\theta}) d \boldsymbol{\theta}\ .
\end{equation}
With the assumption that the different data are independent, the overall likelihood can be written as the product of the likelihood for each individual data:
\begin{equation}
    \label{eq:likelihood}
    p(\mathbf{D} \mid \boldsymbol{\theta})=\prod_{i=1}^n p\left(D_i \mid \boldsymbol{\theta}\right)\ .
\end{equation}

In this work, the data \(\mathbf{D}\) consists of four parts: the NM properties, the HIC flow pressure, the M-R relations of NSs and the tidal deformability of GW170817. More specifically, the NM properties entering the analysis are the six saturation quantities used in Eq.~\eqref{eq:PNM}: the saturation density \(n_0\), the binding energy per nucleon \(E(n_0)\), the incompressibility \(K(n_0)\), the symmetry energy \(E_{\rm sym}(n_0)\), its slope \(L(n_0)\) and the Dirac effective mass ratio \(M_N^*/M_N\). The pressure of symmetric NM in the range $(2-4.6)n_{\rm ref}$ extracted from the $\rm Au+Au$ flow data~\cite{Danielewicz:2002pu} enters as a penalty on the excursion of $P(n)$ outside the band. Here and below, curve locations are quoted in units of the common reference density $n_{\rm ref}=0.16$~fm$^{-3}$, whereas $n_0$ denotes the saturation density predicted by the model. For the M-R relations we use the published inferences of Refs.~\cite{Salmi:2024aum,Choudhury:2024xbk,Mauviard:2025dmd,Doroshenko:2022nwp}, i.e., PSR J0740+6620, PSR J0437$-$4715, PSR J0614$-$3329 and HESS J1731$-$347. Each of them enters Eq.~\eqref{eq:likelihood} as a profile likelihood built from split-normal (asymmetric Gaussian) approximations of its marginal mass and radius posteriors and maximized along the stable M-R sequence of the model, which gives the score in Tab.~\ref{tab:loss} defined by Eq.~\eqref{eq:chiMR}. The tidal information of GW170817 enters through the public low-spin posterior samples~\cite{LIGOScientific:2018hze,LIGOScientific:2018cki}, re-weighted along the $\Lambda(M)$ relation of each EoS as detailed in App.~\ref{app:BJA}; it is not reducible to a single Gaussian constraint on $\Lambda_{1.4}$. Within the BJA framework discussed in detail in App.~\ref{app:BJA}, NM properties and NS observations are incorporated into a unified likelihood~\cite{Guo:2023mhf}, allowing the model parameters to be constrained in a unified, if approximate, manner.

Because the M-R inferences of the two most compact objects are also the most model dependent ones, we do not impose them all at once but consider four fitting scenarios, summarized in Tab.~\ref{tab:sets}: GQHD1 uses only PSR J0740+6620 and PSR J0437$-$4715, i.e., the $2M_\odot$ support together with the radius of a $\sim1.4M_\odot$ star; GQHD2 adds PSR J0614$-$3329; GQHD3 adds HESS J1731$-$347 instead; GQHD4 includes all four objects. All four sets share the same NM targets, the same HIC flow penalty, and the same numerical and stability screening, and are finally refined against the GW170817 tidal posterior. For each set we monitor the total objective $\mathcal{L}_{\rm tot}=-2\,{\rm BJA}+{\rm const}$, i.e., the sum of the $\chi^2$-type terms, defined in Eq.~\eqref{eq:Ltot} and decomposed in Tab.~\ref{tab:loss} of App.~\ref{app:BJA}; since the M-R data entering the four sets are different, the values of $\mathcal{L}_{\rm tot}$ are only comparable within a set, not across sets. The resulting sets are local optima obtained by a continuous local search around the best candidates of a preceding global scan, and should be regarded as representative solutions rather than as a proof of global optimality in the 21-dimensional parameter space.

\begin{table}[htbp]
\centering
\caption{The M-R inferences entering the likelihood of each parameter set, and the total objective $\mathcal{L}_{\rm tot}$ before and after the final refinement against the GW170817 tidal posterior (same objective definition within a set). The NM properties and the HIC flow pressure enter all four sets. Values of $\mathcal{L}_{\rm tot}$ are not comparable across sets since the data differ.}
\label{tab:sets}
\begin{tabular}{lcccccc}
\hline\hline
Set & J0740 & J0437 & J0614 & HESS J1731 & $\mathcal{L}_{\rm tot}^{\rm before}$ & $\mathcal{L}_{\rm tot}^{\rm after}$ \\
\hline
GQHD1 & $\checkmark$ & $\checkmark$ & & & 31.16 & 31.10 \\
GQHD2 & $\checkmark$ & $\checkmark$ & $\checkmark$ & & 32.97 & 32.47 \\
GQHD3 & $\checkmark$ & $\checkmark$ & & $\checkmark$ & 34.79 & 34.60 \\
GQHD4 & $\checkmark$ & $\checkmark$ & $\checkmark$ & $\checkmark$ & 36.32 & 34.80 \\
\hline\hline
\end{tabular}

\end{table}

\section{Nuclear matter properties and neutron star structure}

By using the BJA discussed above, we perform the assessment of model~\eqref{eq:Lag}. For comparison, we also consider several widely used Walecka-type models, including TM1~\cite{Sugahara:1993wz}, NL3~\cite{Lalazissis:1996rd} and FSU-\(\delta 6.7\)~\cite{Li:2022okx}. The reference models are taken with their published parameters and are not refitted; their EoSs, crust matching and stellar structure are computed with exactly the same numerical pipeline as the GQHD sets. In all figures the same line style is used for a given model. Two features of the numerical treatment are relevant for the numbers below. First, the crust is attached to the core by the formal Maxwell construction of Sec.~II, whose crossing lies at $n\simeq1\times10^{-3}$~fm$^{-3}$ for all seven EoSs; the uniform $npe$ matter therefore extends over the density range that is usually described by an inner crust, and the radii and tidal deformabilities quoted below carry a corresponding, unquantified systematic uncertainty. It is common to the seven EoSs, which share the same prescription, and therefore affects the entries of Tab.~\ref{tab:NS} as a whole rather than their differences. Second, the thermodynamic stability and the causality of the EoSs are verified on the density grids and compositions actually computed, rather than established for arbitrary density and composition.

\subsection{NM and NS properties}

\begin{figure}[htbp]
    \centering
    \includegraphics[width=1\linewidth]{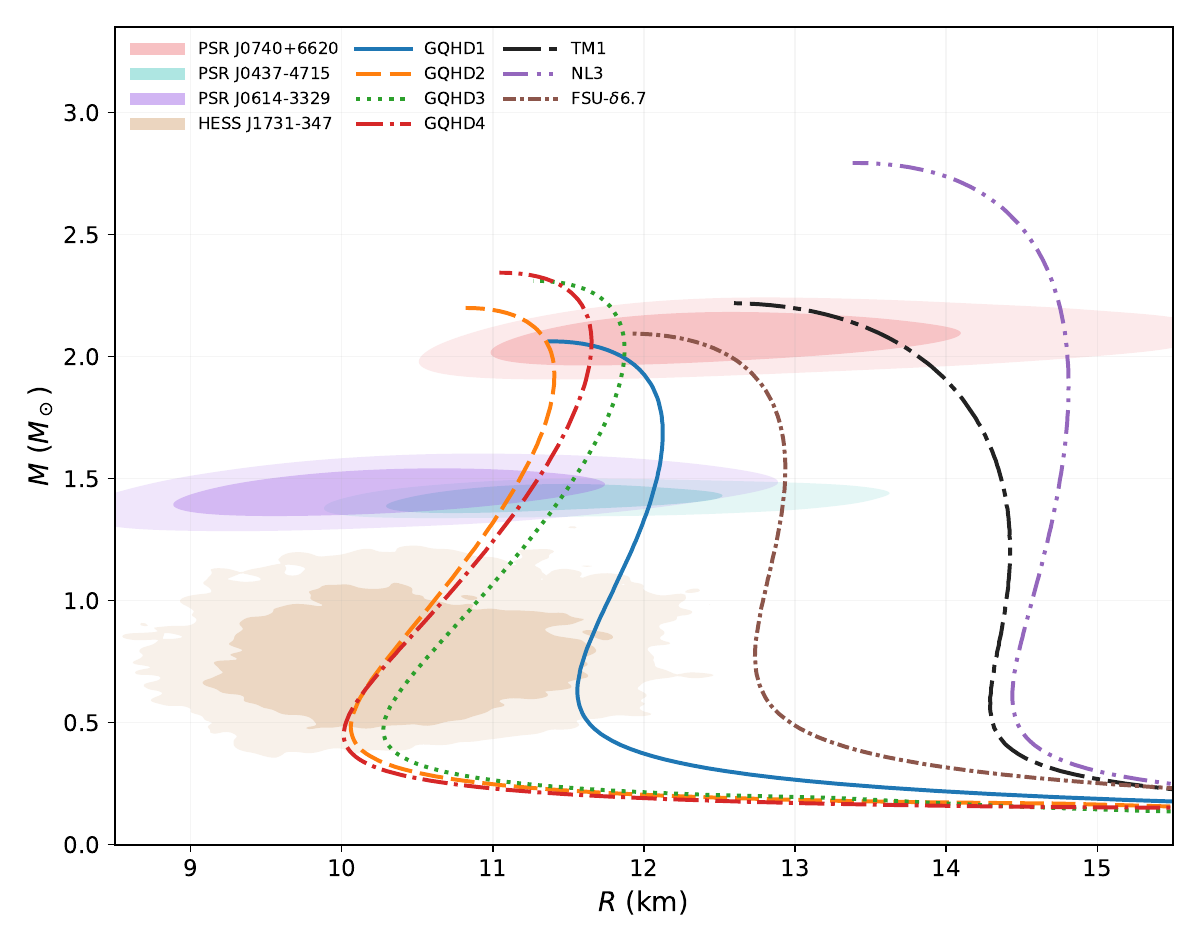}
    \caption{The M-R relation of TM1~\cite{Sugahara:1993wz}, NL3~\cite{Lalazissis:1996rd}, FSU-\(\delta 6.7\)~\cite{Li:2022okx} and GQHD1--GQHD4. The shaded regions are the two-dimensional M-R posteriors of PSR J0740+6620~\cite{Salmi:2024aum} (68.3\% and 95.4\%), PSR J0437$-$4715~\cite{Choudhury:2024xbk}, PSR J0614$-$3329~\cite{Mauviard:2025dmd} and HESS J1731$-$347~\cite{Doroshenko:2022nwp} (approximately 68\% and 95\%, darker and lighter shades). All four objects are shown, but only the subset listed in Tab.~\ref{tab:sets} enters the likelihood of a given set. }
    \label{fig:MROBE}
\end{figure}

From Tab.~\ref{tab:NM} and Fig.~\ref{fig:MROBE} one can clearly observe that, the general model~\eqref{eq:Lag} can accommodate the constraints from experimental data and astro-observations imposed on it, with the residual tensions quantified below. This observation results to a conclusion that the pure-hadronic model can yield the EoSs that reproduce the observed astrophysical M-R ranges while keeping the NM properties close to their empirical ranges, so that the present data do not, by themselves, compel the introduction of exotic DoFs, such as a deconfinement phase transition.

If one were focus on the NM properties, all the considered models roughly reproduce the empirical values, as shown in Tab.~\ref{tab:NM}; the largest deviations are the incompressibility of GQHD2--GQHD4, $K(n_0)=(303-330)$~MeV, above the empirical interval $230\pm30$~MeV, the binding energy of GQHD4, $-14.0$~MeV, and the symmetry energy and its slope of TM1 and NL3. However, the M-R relation predicted by the Walecka-type models fail to cover all credible astrophysical regions as shown in Fig.~\ref{fig:MROBE}, indicating the residual tension between the NM systematics and the astrophysical data. In contrast, the GQHD considered, with parameter sets GQHD1--GQHD4, can reproduce the M-R inferences imposed on them (Tab.~\ref{tab:sets}). GQHD1, constrained by PSR J0740+6620 and PSR J0437$-$4715 only, gives $R_{1.4}=12.0$~km and $M_{\max}=2.06M_\odot$, its M-R curve overlapping the credible regions of both objects. Once PSR J0614$-$3329 or HESS J1731$-$347 is included, the M-R curves move to the left by $(0.6-1.0)$~km at $1.4M_\odot$, $R_{1.4}=(11.1-11.4)$~km, and at the same time the maximum mass increases to $(2.20-2.34)M_\odot$; in particular, GQHD3 and GQHD4 pass through the HESS J1731$-$347 region at $M\simeq0.77M_\odot$ and $R\simeq10.4$~km while still supporting PSR J0740+6620 (Tab.~\ref{tab:NS}). The values of the parameters in sets GQHD1--GQHD4 are given in Tab.~\ref{tab:Par} in App.~B.

It is instructive to see how the solutions respond to the data. The four parameter sets share the same $21$-dimensional structure, yet the addition of a single compact object changes the solution qualitatively: the incompressibility moves from $K(n_0)=238$~MeV in GQHD1 to $(303-330)$~MeV in GQHD2--GQHD4, the slope $L(n_0)$ from $43$ to $(62-71)$~MeV, and several of the meson couplings change by an order of magnitude (Tab.~\ref{tab:Par}). This shows that, for the selected local solutions, the joint nuclear and astrophysical constraints are mutually restrictive and discriminate within the model space; a quantitative statement about how strongly the data constrain the parameter space requires a posterior-volume and model-complexity analysis, which is beyond the present work. Indeed, the present Eq.~\eqref{eq:BJA} only measures how well a model faces the data, but does not account for the model complexity which we leave for the future discussion.

From Fig.~\ref{fig:MROBE} one may find that the size of the NS with intermediate mass, $\approx (1.3-1.5)M_\odot$, is significant for revealing the constituents of NM. A smaller radius, like PSR J0614–3329, may be in tension to most of the pure hadronic description. To obtain a smaller radius, the EoS should be soft, equivalently, a smaller SV to be discussed later, at intermediate density. Consequently, the pressure in this density range must be low. However, to predict a larger maximum NS mass, the EoS should be stiff. In a word, to consistent with the observations, the EoS should be soft at intermediate density but stiff high density relevant to the cores of massive stars.

\begin{widetext}
\begin{table*}[htbp]
\centering
\caption{Nuclear matter properties at saturation density. $n_0$ is saturation density predicted by the model, $E(n_0)$ the binding energy per nucleon, $K(n_0)=9\,dP/dn|_{n_0}=9n_0^2\,d^2(E/A)/dn^2|_{n_0}$, $L(n_0)=3n_0\,dE_{\rm sym}/dn$, $M_N^*(n_0)/M_N$ the Dirac effective mass ratio at saturation, and $P(2n_0)$ the symmetric-matter pressure at $2n_0$. Energies and pressures are in MeV and MeV~fm$^{-3}$. The last row gives the total objective $\mathcal{L}_{\rm tot}=-2\,{\rm BJA}+{\rm const}$ of Tab.~\ref{tab:sets}, with BJA defined in Eq.~\eqref{eq:BJA}; lower is better within a set, and the values are not comparable across sets or defined for the reference models.
}

\label{tab:NM}
{
\begin{tabular}{ccccccccc}
\hline\hline
& Empirical & TM1 & NL3 & FSU-\(\delta 6.7\) & GQHD1 & GQHD2 & GQHD3 & GQHD4 \\
\hline
$E(n_0)$ & $ -16.0 \pm 2.0 $~\cite{Dutra:2014qga} & $-16.3$ & $-16.2$ & $-16.3$ & $-16.0$ & $-15.0$ & $-15.1$ & $-14.0$ \\
$n_0$ & $ 0.155 \pm 0.012 $~\cite{Dutra:2014qga} & 0.145 & 0.148 & 0.148 & 0.159 & 0.156 & 0.157 & 0.159 \\
$E_{\mathrm{sym}}(n_0)$ & $ 31 \pm 4 $~\cite{Oertel:2016bki,Lattimer:2012xj} & 36.9 & 37.4 & 32.7 & 30.3 & 32.9 & 31.3 & 31.5 \\
$L(n_0)$ & $ 60 \pm 30 $~\cite{Lattimer:2012xj,Danielewicz:2013upa,Li:2013ola} & 111 & 119 & 53.5 & 43.3 & 67.7 & 70.9 & 62.4 \\
$K(n_0)$ & $ 240 \pm 80$~\cite{Dutra:2012mb,Stone:2014wza} & 281 & 272 & 229 & 238 & 303 & 330 & 312 \\
$M_N^*(n_0)/M_N$ & --- & 0.634 & 0.595 & 0.610 & 0.606 & 0.706 & 0.684 & 0.699 \\
$P(2n_0)$ & --- & 16.3 & 28.9 & 13.5 & 15.4 & 18.0 & 20.9 & 20.5 \\
$m_\sigma$ & $ 480 \pm 80 $~\cite{ParticleDataGroup:2024cfk} & 511 & 508 & 492 & 534 & 474 & 481 & 495 \\
$\mathcal{L}_{\rm tot}$ & --- & --- & --- & --- & 31.10 & 32.47 & 34.60 & 34.80 \\
\hline\hline
\end{tabular}

}
\end{table*}

\begin{table*}[htbp]
\centering
\caption{NS properties obtained from the seven EoSs with the common crust treatment: the maximum mass, the radii of $1.4M_\odot$ and $2.0M_\odot$ stars, and the dimensionless tidal deformability of a $1.4M_\odot$ star. Values are computed in this work and are not transcribed from the literature.}
\label{tab:NS}
\begin{tabular}{cccccccc}
\hline\hline
 & TM1 & NL3 & FSU-$\delta6.7$ & GQHD1 & GQHD2 & GQHD3 & GQHD4 \\
\hline
$M_{\max}$ ($M_\odot$) & 2.22 & 2.79 & 2.09 & 2.06 & 2.20 & 2.31 & 2.34 \\
$R_{1.4}$ (km) & 14.40 & 14.73 & 12.92 & 12.04 & 11.09 & 11.43 & 11.19 \\
$R_{2.0}$ (km) & 13.82 & 14.80 & 12.57 & 11.85 & 11.39 & 11.87 & 11.65 \\
$\Lambda_{1.4}$ & 1175 & 1384 & 603 & 419 & 281 & 349 & 307 \\
\hline\hline
\end{tabular}

\end{table*}
\end{widetext}

\begin{figure}[htbp]
    \centering
    \includegraphics[width=1\linewidth]{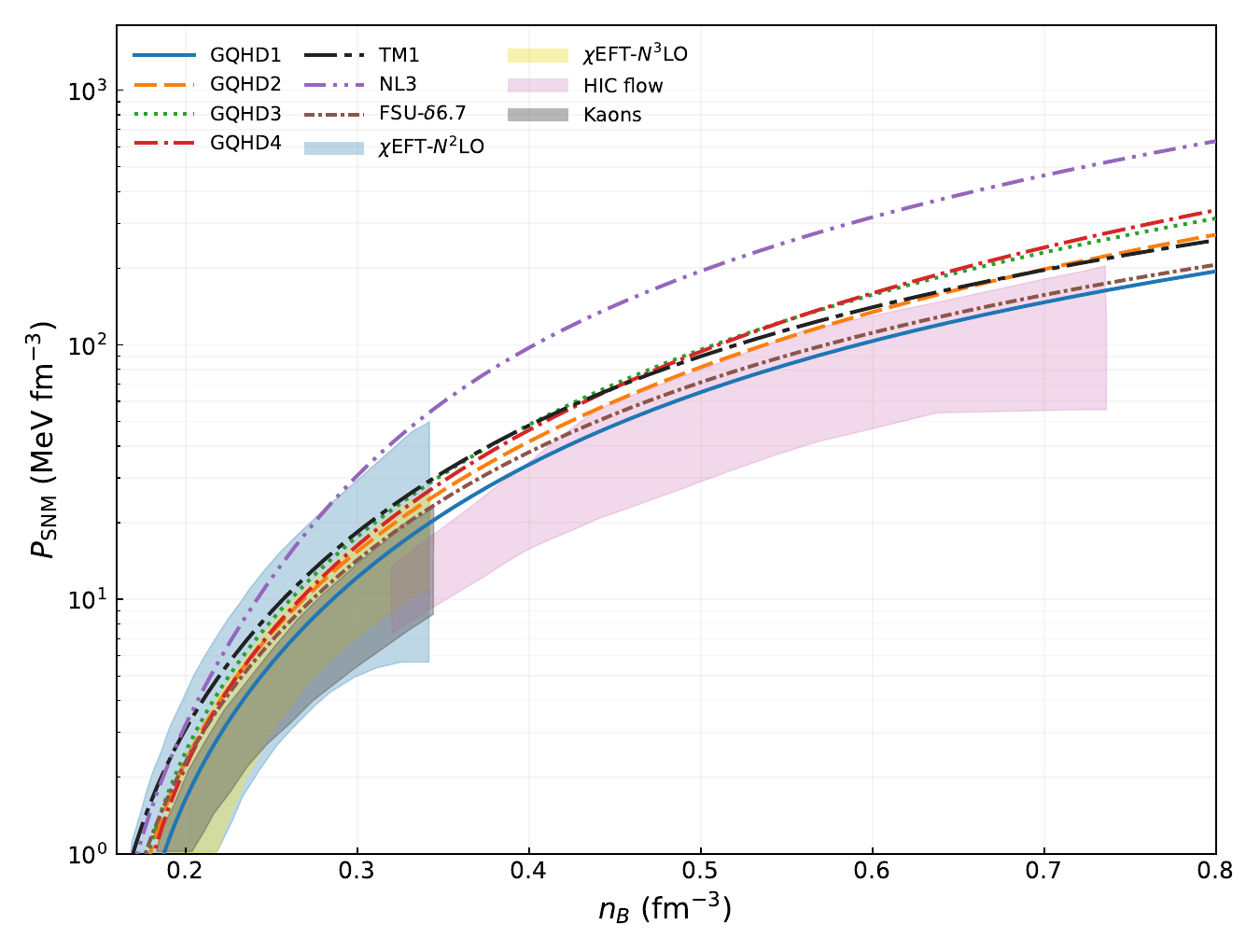}
    \caption{ Pressure as a function of density for symmetric NM. HIC constraint is from $Au+Au$ flow-data analysis~\cite{Danielewicz:2002pu}, $\chi EFT-N^2LO$ and $\chi EFT-N^3LO$ are from chiral nuclear force~\cite{Huth:2021bsp}, Kaons constraint is from~\cite{Lynch:2009vc,Fuchs:2005zg}. Only the HIC flow band enters the BJA; the other regions are shown for comparison and are digitized closed contours without an assigned confidence level.}
    \label{fig:Prho}
\end{figure}

Fig.~\ref{fig:Prho} shows the pressure of symmetric NM together with the constraints from the $\rm Au+Au$ flow data~\cite{Danielewicz:2002pu}, from chiral EFT~\cite{Huth:2021bsp} and from kaon production~\cite{Lynch:2009vc,Fuchs:2005zg}. The flow band is part of the calibration of GQHD1--GQHD4 through the penalty of Eq.~\eqref{eq:PHIC}; the chiral EFT and the kaon regions enter nowhere in the fit, and the reference models are not refitted at all.

GQHD1 and FSU-$\delta6.7$ lie inside the flow band above $\sim2.5n_{\rm ref}$, GQHD2 runs along its upper edge, GQHD3, GQHD4 and TM1 lie $(10-35)\%$ above the upper edge, and NL3 exceeds it by more than a factor of two; near the low-density end, $n\lesssim2.5n_{\rm ref}$, all the models are slightly above it. The corresponding flow penalties are $0.18$, $1.56$, $4.20$ and $3.46$ for GQHD1--GQHD4 in the units of $\mathcal{L}_{\rm tot}$ (Tab.~\ref{tab:loss}), the two largest values belonging to the sets that reproduce the M-R inference of HESS J1731$-$347. At low density, chiral EFT at N$^3$LO and the kaon region disagree with NL3, while the N$^2$LO region is too wide to discriminate between the models considered.

The $\beta$-equilibrated EoSs of GQHD2--GQHD4 are stiffer than that of FSU-$\delta6.7$ at high density, and their maximum masses, $(2.20-2.34)M_\odot$, are correspondingly larger, in agreement with the observations of massive NSs, as illustrated in Fig.~\ref{fig:MROBE}.

\begin{figure*}[htbp]
    \centering
    \includegraphics[width=0.95\linewidth]{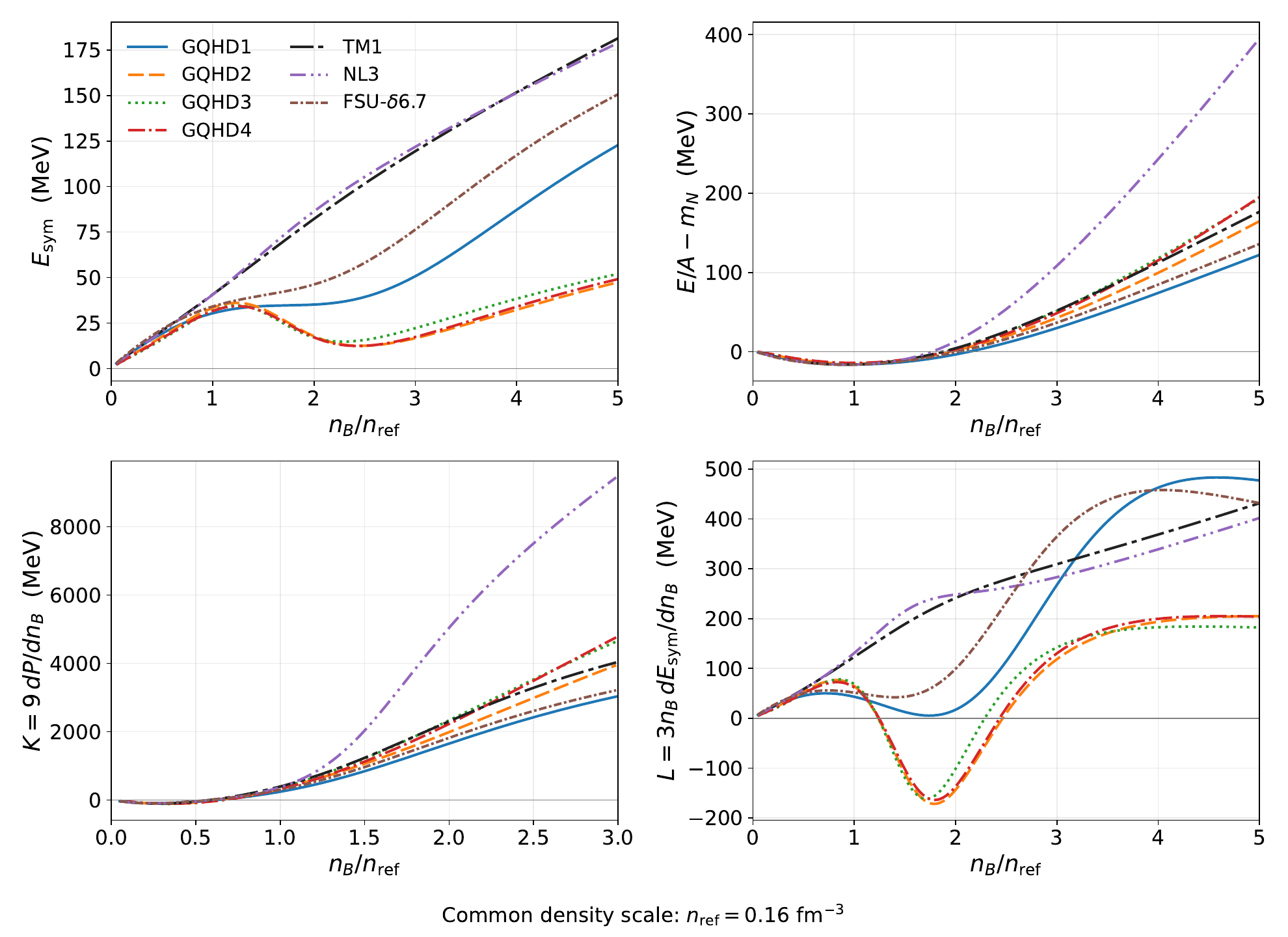}
    \caption{Comparison of the density dependence of nuclear matter properties in various of models. The symmetry energy (upper left), the binding energy per nucleon of symmetric matter (upper right), the incompressibility $K(n)=9\,dP/dn$ (lower left) and the symmetry-energy slope $L(n)=3n\,dE_{\rm sym}/dn$ (lower right); the density is scaled by a common reference $n_{\rm ref}=0.16$~fm$^{-3}$.
    }
    \label{fig:nuclearMatterProperties}
\end{figure*}

Fig.~\ref{fig:nuclearMatterProperties} compares the density dependence of several nuclear matter properties predicted by different RMF models.
The binding energy per nucleon exhibits a minimum around the saturation density, reflecting the saturation mechanism of symmetric nuclear matter.
While the models behave similarly near saturation, they diverge rapidly at suprasaturation densities.
The symmetry energy and its slope display a sizable model dependence. Explicitly, NL3 and TM1 generally yield a stiffer high-density behavior, whereas FSU-$\delta 6.7$ and GQHD1 remain softer.
In particular, GQHD2--GQHD4 behave qualitatively differently: their symmetry energy reaches a local maximum of $\simeq(34-36)$~MeV around $1.25n_{\rm ref}$, decreases to a minimum of $(12-15)$~MeV around $(2.3-2.5)n_{\rm ref}$, and only then rises again. Correspondingly, they develop a negative symmetry-energy slope in the low-to-intermediate density region, $L(n)<0$ between $\simeq1.25n_{\rm ref}$ and $\simeq(2.3-2.5)n_{\rm ref}$, reaching $\simeq-(160-170)$~MeV near $1.8n_{\rm ref}$, whereas for GQHD1 the slope only dips to $\simeq+5$~MeV near $1.8n_{\rm ref}$ without changing sign. The symmetry energy stays positive throughout for all sets. This softening strongly softens $\beta$-equilibrated matter and thus leads to more compact stars in the mass-radius relation.
The incompressibility, defined at any density by $K(n)=9\,dP/dn$, which reduces to $9n_0^2\,d^2(E/A)/dn^2$ at saturation where $P$ vanishes, further illustrates the different high-density stiffness of the EoSs, with NL3 rising most rapidly above saturation density: it reaches $5.0$~GeV at $2n_{\rm ref}$, twice the value of any other model. Among the remaining ones GQHD3, GQHD4 and TM1 are comparable near $2n_{\rm ref}$, $(2.2-2.3)$~GeV, while above $\simeq2.5n_{\rm ref}$ GQHD3 and GQHD4 become the largest, $(4.7-4.8)$~GeV at $3n_{\rm ref}$ against $4.0$~GeV for TM1; GQHD1 and FSU-$\delta6.7$ stay the softest at all these densities. Below $\simeq(0.6-0.7)n_{\rm ref}$, $K(n)$ turns slightly negative for all the models, with a minimum of about $-100$~MeV, the uniform symmetric matter being mechanically unstable there.
Consistently, as shown in Fig.~\ref{fig:MROBE}, the M-R constraints favor the softer EoSs in the radius-sensitive density region, since they produce more compact intermediate-mass NSs.

\begin{figure*}[htbp]
    \centering
    \includegraphics[width=0.95\linewidth]{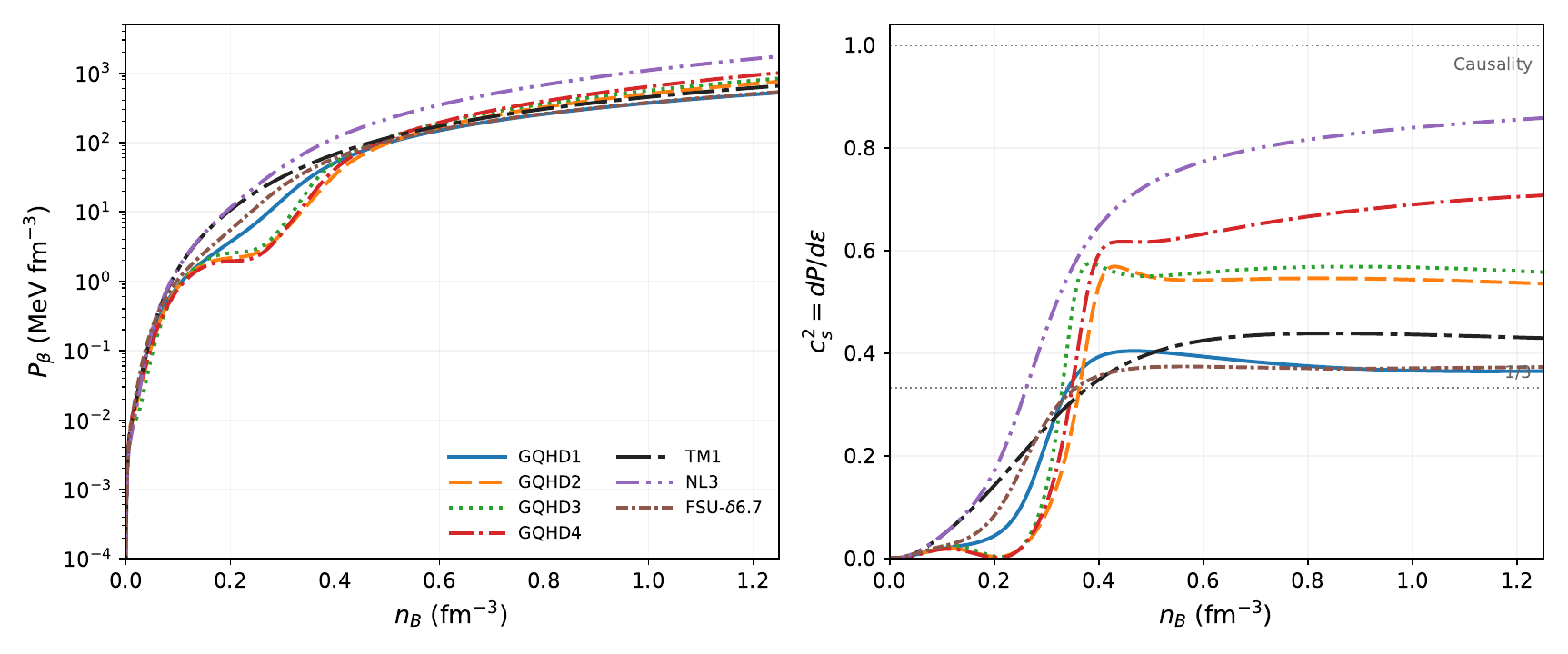}
    \caption{Left: Density dependence of the pressure in beta-equilibrated matter after matching the low-density BPS crust to the RMF core EOS. Right: the density dependence of sound velocity $v_s^2=dP/d\varepsilon$ (labelled $c_s^2$ in the figure) of the same matter; the dotted lines mark the conformal value $1/3$ and the causal limit. The sound velocity is evaluated within the uniform core and does not include the crust-core junction.}
    \label{fig:Pbeta}
\end{figure*}

The left panel of Fig.~\ref{fig:Pbeta} shows the pressure of $\beta$-equilibrated matter after matching the RMF core EoS to the BPS crust at low densities.
For all parameter sets the pressure increases monotonically with density, and the relative stiffness at high density follows the same ordering as in the symmetric-matter analysis, with NL3 giving the largest pressures.
For GQHD2--GQHD4, the pressure is strongly suppressed in the low-to-intermediate density region owing to the negative symmetry-energy slope: it develops a nearly flat shoulder at $n\simeq(0.15-0.24)$~fm$^{-3}$, $P\simeq(2-3)$~MeV~fm$^{-3}$, and then rises steeply to join the other curves near $0.45$~fm$^{-3}$; a milder version of the same shoulder is visible for GQHD1. This is precisely the density window probed by the radii of $(1.0-1.5)M_\odot$ stars.

\subsection{Sound velocity}
\label{sec:SV}

Let us consider the SV \(v_s^2 = dP/d\varepsilon\) (in units of $c=1$; denoted $c_s^2$ in the figures) of $\beta$-equilibrated matter, which measures the local stiffness of the EoS. The results of the four GQHD sets and of the Walecka-type models TM1, NL3 and FSU-$\delta6.7$ are shown in the right panel of Fig.~\ref{fig:Pbeta}; as in Fig.~\ref{fig:nuclearMatterProperties}, densities are quoted below in units of $n_{\rm ref}=0.16$~fm$^{-3}$.

The three Walecka-type models behave in the familiar way. Their SV is already sizable at saturation, $v_s^2\simeq(0.05-0.11)$ at $n_{\rm ref}$, grows smoothly and almost monotonically, crosses the conformal value $1/3$ at $(1.6-2.4)n_{\rm ref}$ and saturates at high density, reaching $v_s^2\simeq0.37$ (FSU-$\delta6.7$), $0.43$ (TM1) and $0.86$ (NL3) at $n=1.25$~fm$^{-3}$. The rise is spread over the whole density range of the star, roughly from $0.6n_{\rm ref}$ to $3n_{\rm ref}$, and no pronounced maximum develops: TM1 and FSU-$\delta6.7$ have only shallow maxima, their decrease after the highest value being below $0.01$. In this smooth saturation the vector repulsion gradually takes over the scalar attraction; the three parametrizations lack both the extended soft window and the abrupt rise described next.

The GQHD sets are qualitatively different, as shown in the right panel of Fig.~\ref{fig:Pbeta}. All four develop a two-stage density dependence consisting of a soft window, extended and very pronounced for GQHD2--GQHD4 and milder for GQHD1, followed by an abrupt rise; GQHD1--GQHD3 then pass through a local maximum, modest in GQHD1, whereas GQHD4 levels off. We refer to this non-monotonic behavior as the peak structure. For GQHD2--GQHD4, the sets accommodating PSR J0614$-$3329 and/or HESS J1731$-$347, the SV stays at the level $v_s^2\sim10^{-2}$ up to $\simeq1.5n_{\rm ref}$, an order of magnitude below the Walecka-type models at the same density, and then jumps between $\simeq1.7n_{\rm ref}$ and $\simeq2.5n_{\rm ref}$ to $v_s^2\simeq(0.5-0.6)$; the steepest slope, $dv_s^2/d(n/n_{\rm ref})\simeq(1.0-1.4)$, is two to seven times larger than in the Walecka-type models. GQHD2 and GQHD3 pass through a maximum, $v_s^2=0.57$ at $2.7n_{\rm ref}$ and $0.58$ at $2.4n_{\rm ref}$ respectively, and then decrease slightly to a plateau of $\simeq(0.54-0.57)$, whereas for GQHD4 the abrupt rise is followed by a slow, nearly monotonic further increase to $0.71$ at $1.25$~fm$^{-3}$. GQHD1, which is constrained by the two heavier pulsars only, shows a milder version of the same shape: $v_s^2\simeq0.03$ at $n_{\rm ref}$, a rise between $\simeq1.2n_{\rm ref}$ and $\simeq2.3n_{\rm ref}$, a broad maximum of $0.40$ at $\simeq2.9n_{\rm ref}$ and a decrease to $0.37$ at the highest density. All seven EoSs remain causal up to the central density of their maximum-mass configurations, the largest value, $v_s^2=0.79$, being reached by NL3.

Two features of this structure deserve emphasis. First, the maximum of $v_s^2$ in GQHD1--GQHD3 is located at $(2.4-2.9)n_{\rm ref}$, i.e., at about the central density of a $1.4M_\odot$ star in these sets, $(2.6-3.0)n_{\rm ref}$, and far below the central density of their maximum-mass stars, $(5.7-6.3)n_{\rm ref}$. The soft window and most of the abrupt rise are therefore sampled by the intermediate-mass stars: the low pressure in this density range is what produces the small radii $R_{1.4}=(11.1-11.4)$~km of GQHD2--GQHD4, whereas in the Walecka-type models, whose SV has already reached $v_s^2\simeq(0.26-0.35)$ at the center of a $1.4M_\odot$ star, the intermediate-mass stars are large. Second, beyond the rise the SV settles at $v_s^2\simeq(0.55-0.7)$, well above the conformal value $1/3$, and this high-density stiffness is what supports the maximum masses $(2.2-2.3)M_\odot$ of GQHD2--GQHD4. The two requirements from the observations, compactness at intermediate mass and a large maximum mass, are thus met by one and the same EoS. In this sense the non-monotonic SV partitions the EoS into a soft segment at low-to-intermediate density and a stiff segment at high density, and the sequential measurement of NSs with different masses becomes a direct probe of this partition.

A SV of this shape, strongly suppressed at $(1-1.5)n_{\rm ref}$, rising rapidly and exceeding $1/3$ at a few $n_{\rm ref}$ before leveling off, is usually taken as a fingerprint of a change of the DoFs, for example of quarkyonic matter~\cite{McLerran:2018hbz} or of a crossover to quark matter~\cite{Annala:2019puf,Altiparmak:2022bke}, and, as Fig.~\ref{fig:Pbeta} shows, it is not produced by the standard Walecka-type parametrizations. Here it emerges within a purely nucleonic RMF description, without any change of DoFs, as the response of the general Lagrangian to the joint nuclear and astrophysical constraints: the soft-to-stiff structure is present in all four sets, and it is most pronounced in those that reproduce the smallest objects. Non-monotonic sound speeds have been obtained before in hadronic effective theories with dilaton or topology-change dynamics~\cite{Zhang:2024sju,Zhang:2024iye,Paeng:2017qvp,Ma:2018jze}, and the present result reaffirms our previous conclusion that a peak of the SV can arise in pure hadronic models; what is new here is that the structure appears in a conventional, polynomial Walecka-type Lagrangian, together with compact intermediate-mass radii and $2M_\odot$ support, and it therefore does not by itself identify a change of the microscopic DoFs. In the present sets it appears as an abrupt soft-to-stiff transition followed by a broad maximum or plateau rather than as a narrow peak. We stress that the structure is a property of the full set of dimension-4 operators acting together and we do not attribute it to any individual coupling here; disentangling the role of the different operators requires a dedicated analysis, which will be presented separately.

Finally, the SV also explains the failure of the Walecka-type models in Fig.~\ref{fig:MROBE}. NL3, with the largest SV at all densities, gives the stiffest EoS and consequently both the largest maximum mass and the largest radii; TM1 and FSU-$\delta6.7$ have a moderate SV at high density, close to the conformal value, but their SV is already large at $(1-2)n_{\rm ref}$, in line with their stiff symmetry energy (Fig.~\ref{fig:nuclearMatterProperties}), so that the radii of the intermediate-mass stars remain too large to be compatible with PSR J0614$-$3329 and HESS J1731$-$347.

\subsection{Tidal deformability}

\begin{figure}[htbp]
    \centering
    \includegraphics[width=1\linewidth]{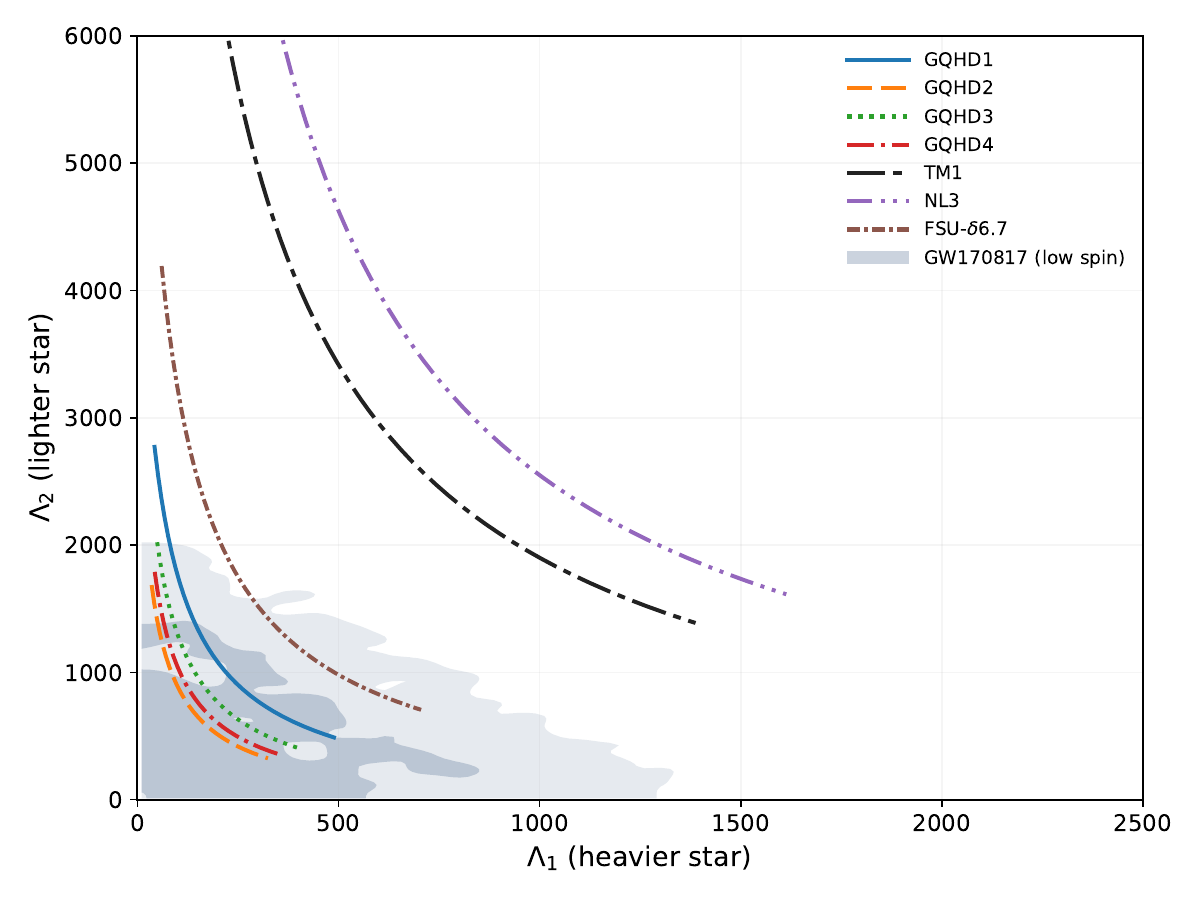}
    \caption{The tidal deformabilities $\Lambda_1$ and $\Lambda_2$ of the heavier and lighter components of GW170817, compared with the low-spin posterior of Ref.~\cite{LIGOScientific:2018hze} (darker and lighter shades: approximately 68\% and 95\%). The model curves are obtained for the fixed source-frame chirp mass $\mathcal{M}=1.186M_\odot$ by varying the mass ratio; they are display slices and are not the quantity integrated in the likelihood.}
    \label{fig:Binary}
\end{figure}

\begin{figure}[htbp]
    \centering
    \includegraphics[width=1\linewidth]{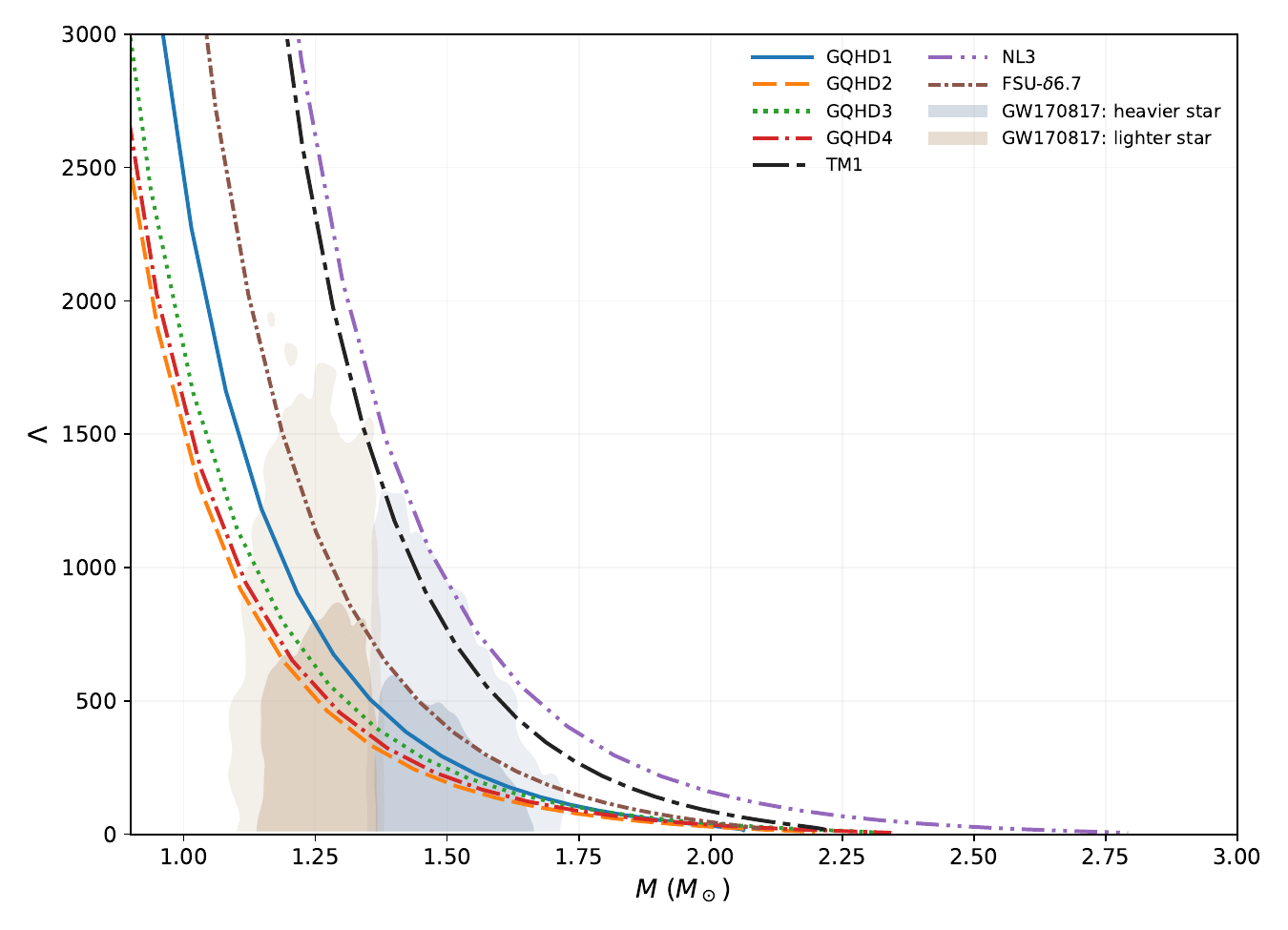}
    \caption{The tidal deformability as a function of the NS mass, obtained from the seven EoSs, compared with the projections of the GW170817 low-spin posterior~\cite{LIGOScientific:2018hze} onto the mass and tidal deformability of the heavier and of the lighter component. The shaded regions are projections of a joint posterior, not EoS-independent confidence bands.}
    \label{fig:MLambda}
\end{figure}

We finally consider another independent constraint from astrophysics, the tidal deformability (TD). Here, we consider the TD inferred from GW170817 which was firstly estimated to be $\Lambda_{1.4} \leq 800$~\cite{LIGOScientific:2017vwq} and then stringent to $\Lambda_{1.4}=190^{+390}_{-120}$~\cite{LIGOScientific:2018cki} for NS with mass $1.4M_\odot$. We refer to the later as the constraint. In the present analysis the GW information enters the likelihood through the joint posterior of the two tidal deformabilities as described in Sec.~III, and the published $\Lambda_{1.4}$ interval is used only as a cross-check. In addition, the correlations between TD of the heavier NS ($\Lambda_1$) and the lighter NS ($\Lambda_2$) of the binary system are also plotted~\cite{LIGOScientific:2018cki,LIGOScientific:2018hze,ligo_p1800115,Kalita:2025keo} in Fig.~\ref{fig:Binary}, together with the projections of the joint posterior onto the mass and TD of each component in Fig.~\ref{fig:MLambda}.

From Tab.~\ref{tab:NS} and Fig.~\ref{fig:MLambda} one can conclude that, the ordering of $\Lambda(M)$ follows that of the radii: the GQHD predicts $\Lambda_{1.4} = 419$, $281$, $349$ and $307$ for GQHD1--GQHD4, compared with $603$ for FSU-$\delta$6.7, $1175$ for TM1 and $1384$ for NL3. All four GQHD sets are inside the $\Lambda_{1.4}=190^{+390}_{-120}$ interval, and their $\Lambda_1$-$\Lambda_2$ curves pass through the high-probability region of the joint posterior in Fig.~\ref{fig:Binary}, GQHD2--GQHD4 lying deepest inside, while FSU-$\delta$6.7 runs along the edge of the outer region and TM1 and NL3 lie outside the shown contours. In Fig.~\ref{fig:MLambda}, the GQHD curves cross the high-probability parts of the posterior projections of both components, FSU-$\delta$6.7 crosses only their outer parts, TM1 grazes the outer region of the heavier component and NL3 passes above both. These projections and the fixed-chirp-mass slice are illustrative displays of the joint posterior and not a test of joint compatibility, which enters only through the integrated likelihood of Sec.~III. The tidal information is therefore consistent with the small radii inferred from the X-ray observations and disfavors the stiff symmetry energy of the traditional parametrizations. 

\section{Conclusion and outlook}

In this work, by using the most general Quantum Hadrodynmics (GQHD) model written up to dimension-4 incorporating the non-strange scalar and vector mesons below $1$~GeV, i.e., \(\sigma, \omega, \rho, a_0\), and \(n, p\), we clarified that the data from nuclear physics experiments and astrophysics observations---the empirical saturation properties, the HIC flow pressure, the M-R inferences of PSR J0740+6620, PSR J0437$-$4715, PSR J0614$-$3329 and HESS J1731$-$347, and the tidal posterior of GW170817---can be accommodated within the present purely hadronic framework, with a residual tension in the incompressibility of the sets that reproduce the most compact objects. Of course, this conclusion does not imply the absence of exotic phases or DoFs but simply and inversely implies that the currently available nuclear and astrophysical data do not require the introduction of exotic DoFs, such as quark matter or quarkyonic matter.

To systematically estimate the deviation from the constraints, we introduced the BJA framework, an approximate joint-likelihood method that enables a quantitative and unified assessment of model performance across different parameter sets by jointly incorporating nuclear and astrophysical constraints. Within this framework, we constructed four parameter sets that differ only in which M-R inferences are imposed. The set constrained by PSR J0740+6620 and PSR J0437$-$4715 alone, GQHD1, gives $R_{1.4}=12.0$~km and $M_{\max}=2.06M_\odot$; the sets that in addition accommodate PSR J0614$-$3329 and/or HESS J1731$-$347 give $R_{1.4}=(11.1-11.4)$~km, $M_{\max}=(2.20-2.34)M_\odot$ and $\Lambda_{1.4}=280-350$, all within the $\Lambda_{1.4}$ interval inferred from GW170817.
 In addition, we found that the SV of the GQHD develops a peak structure which is absent in the Walecka-type models TM1, NL3 and FSU-$\delta6.7$: in the sets that accommodate the most compact objects an extended soft window, $v_s^2\sim10^{-2}$ up to $\sim1.5n_{\rm ref}$, accompanied by a pronounced softening of the symmetry energy between $\sim1.25n_{\rm ref}$ and $\sim(2.3-2.5)n_{\rm ref}$, is followed by an abrupt rise to $v_s^2\simeq(0.5-0.6)$ around $(2-2.5)n_{\rm ref}$ and a broad maximum (GQHD1--GQHD3) or plateau (GQHD4) well above the conformal value. Such a structure, usually regarded as a signature of exotic DoFs, has to our knowledge not been found before in the Walecka-type RMF models. It is a property of the full set of dimension-4 operators acting together; a systematic investigation of the individual effects of the nonlinear and crossed interaction terms on nuclear matter properties and neutron-star observables will be presented in our forthcoming comprehensive study.

Our results indicate that the smaller size of intermediate mass NSs may give a strong constraint on the NS EoS. In such a case, a soft EoS, i.e. a smaller SV, is needed at intermediate density. However, to obtain the massive mass NS around $2M_\odot$, the EoS must be stiff enough above the central density of the intermediate mass of NS. Therefore a soft-to-stiff transition of SV should exist, which is naturally realized in the present GQHD, most prominently in the GQHD2--GQHD4 parameter sets. This implies that the sequential measurement of neutron star mass and radius by the next generation facilities, especially that of the intermediate mass neutron stars, is crucial for diagnosing the pure nucleonic stars.

In the literature, there are several other preconceived composition of compact star matter, such as the quarkyonic matter and quark matter. These structures are widely discussed and results of the models are consistent with observations more or less. Therefore, it is significant to investigate the boundary of these structures, including the present one, to give the precision of the measurement and novel signal to distinguish different structures, and to pin down the location of the hadron-exotic structure if the latter exists.

The last issue we want to address is that the present GQHD is the pure phenomenological one. No full attention has been paid to the fundamental symmetries of QCD, such as chiral symmetry that has been recognized to control the dynamics of hadron processes. This is a another point deserves future exploration.

\sect{Note added}After this paper was submitted, a Bayesian inference analysis found a preference for a strong first-order phase transition whose onset most likely lies above the central density of the maximum-mass NS~\cite{Cao:2026dwk}. This is compatible with the absence of a strong first-order transition inside stable NSs, although it does not by itself establish a purely hadronic composition, and it supports the view that the NM properties up to the cores of the massive NSs can be described by hadronic DoFs.

\sect{Acknowledgments} The work of Y. M. is supported by Jiangsu Funding Program for Excellent Postdoctoral Talent under Grant Number 2025ZB516. Y.~L. M. is supported in part by the National Science Foundation of China (NSFC) under Grant No. 12547104, the National Key R\&D Program of China under Grant No. 2021YFC2202900 and Gusu Talent Innovation Program under Grant No. ZXL2024363.

\newpage

\begin{widetext}

\appendix

\counterwithin{table}{section}
\counterwithin{figure}{section}

\section{Bayesian joint analysis}
\label{app:BJA}

Based on Bayes' theorem, our primary interest lies in the posterior distribution \(p(\theta \mid \mathbf{D})\) with $\theta$ being the parameter space and $\mathbf{D}$ being the distribution of the data. Since the evidence term \(p(\mathbf{D})\) is independent of the model parameters \(\theta\), it can be treated as a normalization constant, thus neglected in parameter inference. Consequently, by using Eq.~(\ref{eq:likelihood}), Eq.~(\ref{eq:Bayes}) can be rewritten as:
\begin{equation}\label{eq:posterior}
p(\theta \mid \mathbf{D}) \propto \prod_{i=1}^n p\left(D_i \mid \boldsymbol{\theta}\right) p(\theta).
\end{equation}
We further assume a Gaussian likelihood for \(p(\mathbf{D} \mid \theta)\), corresponding to the normally distributed observational errors. Equation~\eqref{eq:posterior} holds for all the data listed in Sec.~III. The likelihood factors constructed in the present analysis do not all have the same form: they are Gaussian for the NM targets, split-normal for the M-R inferences, a one-sided quadratic penalty for the HIC flow band and a kernel density estimate for the GW170817 posterior, and the physical requirements enter through the barriers specified below. In the absence of prior preference, uniform priors are adopted for all parameters within their allowed ranges, supplemented by the physical requirements that the EoS be thermodynamically stable, causal, and support a $2M_\odot$ star; parameter sets violating these requirements are rejected. 

To avoid numerical overflow arising from the product of likelihood terms, particularly in regions of high likelihood, we define BJA as:
\begin{equation}\label{eq:BJA}
\rm BJA =ln\left( \prod_{i=1}^n p\left(D_i \mid \boldsymbol{\theta}\right) p(\theta)\right).
\end{equation}

Each factor entering Eq.~\eqref{eq:BJA} is written as $p_i\propto\exp(-f_i/2)$, so that ${\rm BJA}=-\tfrac{1}{2}\sum_i f_i+{\rm const}$, the constant collecting the normalizations, which do not depend on $\boldsymbol{\theta}$. Maximizing the posterior is therefore equivalent to minimizing the $\chi^2$-like objective
\begin{equation}\label{eq:Ltot}
\mathcal{L}_{\rm tot}\equiv-2\,{\rm BJA}+{\rm const}=\sum_i f_i
=\sum_{j\in s}\chi^2_j+\mathcal{P}_{\rm NM}+\mathcal{P}_{\rm HIC}+\mathcal{L}_{\rm GW}+\mathcal{P}_{\rm reg}+\mathcal{P}_{\rm phys}\ ,
\end{equation}
and the four parameter sets are obtained by this minimization rather than by sampling the posterior. Here $s$ is the set of M-R objects imposed in the scenario (Tab.~\ref{tab:sets}): $\chi^2_j$ is the profile score of the M-R inference of the object $j$, $\mathcal{P}_{\rm NM}$ measures the deviation of the saturation properties from the targets of the fit, $\mathcal{P}_{\rm HIC}$ the excursion of the symmetric-matter pressure outside the HIC flow band, $\mathcal{L}_{\rm GW}$ is the GW170817 term, $\mathcal{P}_{\rm reg}$ regularizes the local search and $\mathcal{P}_{\rm phys}$ collects the barriers enforcing the physical requirements. With this convention a Gaussian factor contributes its $\chi^2$, and a smaller $\mathcal{L}_{\rm tot}$ means a larger posterior. The additive constant is independent of $\boldsymbol{\theta}$ and cancels in any comparison made with the same data, so only such differences are meaningful.

The definitions below fix the form of each term. The numerical configuration of the objective---the intervals and scales of the barriers, the density grids, the quadrature settings of $I_{\rm GW}$ and the parameters of the local search---together with the code that evaluates $\mathcal{L}_{\rm tot}$ and the digitized flow band, is collected in the public repository \url{https://github.com/Xiongjiaying-phy/GQHD}.

\emph{M-R inferences.} For each object $j$ the published marginal posteriors of the mass and of the radius are approximated by split normals, specified by the medians $\hat{M}_j$, $\hat{R}_j$ and by the two one-sided credible widths $\sigma^{\pm}_{M_j}$, $\sigma^{\pm}_{R_j}$. The two marginals are treated as independent and their product is maximized along the stable M-R sequence $R_{\boldsymbol\theta}(M)$ of the EoS, which gives the profile score $\chi^2_j$, i.e. $-2\ln$ of the maximized product up to a constant,
\begin{equation}\label{eq:chiMR}
\chi^2_j=\min_{M}\left[\left(\frac{M-\hat{M}_j}{\sigma^{\pm}_{M_j}}\right)^{2}
+\left(\frac{R_{\boldsymbol\theta}(M)-\hat{R}_j}{\sigma^{\pm}_{R_j}}\right)^{2}\right]\ ,
\end{equation}
the upper or the lower width being used according to the sign of the corresponding residual. The minimization over the mass, rather than an evaluation at $\hat{M}_j$, is necessary because the sequence of a given EoS does not always reach the median mass of the object: GQHD1, for instance, has $M_{\max}=2.06M_\odot$, slightly below the median mass of PSR J0740+6620. The adopted values are $(M,R)=(2.073\pm0.069\,M_\odot,\,12.49^{+1.28}_{-0.88}~{\rm km})$ for PSR J0740+6620~\cite{Salmi:2024aum}, $(1.418\pm0.037\,M_\odot,\,11.36^{+0.95}_{-0.63}~{\rm km})$ for PSR J0437$-$4715~\cite{Choudhury:2024xbk}, $(1.44^{+0.06}_{-0.07}\,M_\odot,\,10.29^{+1.01}_{-0.86}~{\rm km})$ for PSR J0614$-$3329~\cite{Mauviard:2025dmd} and $(0.77^{+0.20}_{-0.17}\,M_\odot,\,10.40^{+0.86}_{-0.78}~{\rm km})$ for HESS J1731$-$347~\cite{Doroshenko:2022nwp}. Since the product of the two marginals is not the published two-dimensional posterior, which is correlated, $\chi^2_j$ measures a distance in units of the marginal widths and does not correspond to a credible level of that posterior.

\emph{NM properties.} With $X_a$ denoting the six saturation quantities $(n_0,E(n_0),K(n_0),E_{\rm sym}(n_0),L(n_0),M^*/M)$ predicted by a parameter set,
\begin{equation}\label{eq:PNM}
\mathcal{P}_{\rm NM}=\sum_a\left(\frac{X_a-c_a}{s_a}\right)^{2}+\mathcal{B}\ ,\qquad
\mathcal{B}=w_{\rm b}\sum_a\frac{[\,\ell_a-X_a\,]_+^2+[\,X_a-h_a\,]_+^2}{s_a^{\prime\,2}}\ ,
\end{equation}
where the centres and scales $(c_a,s_a)$ are $(0.155,0.012)$~fm$^{-3}$, $(-16,2)$~MeV, $(240,80)$~MeV, $(31,4)$~MeV, $(60,30)$~MeV and $(0.65,0.15)$, and $[x]_+=\max(x,0)$. The second term is a steep one-sided barrier which vanishes as long as every $X_a$ stays inside the acceptance window $[\ell_a,h_a]$ and grows quadratically outside it, the weight $w_{\rm b}$ being large compared with unity; the windows $[\ell_a,h_a]$, the scales $s'_a$ and $w_{\rm b}$ are given in the repository quoted above. 
This penalty discourages the optimizer from improving the astrophysical fit at the expense of moving nuclear-matter properties outside the prescribed acceptance windows.
These centres and scales are the penalty scales of the fit, chosen to span the ranges allowed by the data; they are not independent experimental $1\sigma$ errors and should not be confused with the empirical determinations quoted, with their references, in Tab.~\ref{tab:NM}. The mass of the $\sigma$ meson is not penalized, being simply bounded as stated in Sec.~II.

\emph{HIC flow band.} With $u=n/n_{\rm ref}$ and $P(u)$ the pressure of symmetric NM, and with $P_{\rm lo}(u)$ and $P_{\rm hi}(u)$ the lower and upper boundaries of the flow band of Ref.~\cite{Danielewicz:2002pu},
\begin{equation}\label{eq:PHIC}
\mathcal{P}_{\rm HIC}=\frac{w}{u_2-u_1}\int_{u_1}^{u_2}r^2(u)\,du\ ,\qquad
r=\frac{[\,\ln(P_{\rm lo}/P)\,]_++[\,\ln(P/P_{\rm hi})\,]_+}{\tfrac{1}{2}\ln(P_{\rm hi}/P_{\rm lo})}\ ,
\end{equation}
with $(u_1,u_2)=(2.0,4.6)$ and $w=10$. The excursion is measured in the logarithm of the pressure, in units of the half-width of the band, and vanishes identically inside it, so that the term penalizes departures from the band without attracting the EoS towards its centre. The weight $w$ is a hyperparameter of the fit and not a statistical weight, and the band is a digitization of Fig.~3 of Ref.~\cite{Danielewicz:2002pu} rather than an original numerical table.

\emph{GW170817.} The tidal term is $\mathcal{L}_{\rm GW}=-2\ln I_{\rm GW}$, where $I_{\rm GW}$ is a kernel density estimate of the released low-spin PhenomPNRT posterior samples in the four dimensions $(\mathcal{M},q,\Lambda_1,\Lambda_2)$, evaluated on the track defined by the $\Lambda(M)$ relation of the EoS and integrated over the detector-frame chirp mass and the mass ratio. The mass prior is taken to be the one of the original parameter estimation, so that the corresponding Jacobian cancels in the event integral, and the tidal deformabilities keep the uniform priors of that analysis. This re-weights the published parameter posterior and is not a re-analysis of the GW strain. The additive normalization of $\ln I_{\rm GW}$ is arbitrary and common to all sets, so the absolute size of $\mathcal{L}_{\rm GW}$ carries no goodness-of-fit information and only its variation between neighbouring parameter sets is meaningful.

\emph{Regularization and physical barriers.} The refinement is a bounded local search around a starting candidate, and $\mathcal{P}_{\rm reg}$ is a quadratic penalty on the displacement from it, which keeps the search in that neighbourhood. Finally, $\mathcal{P}_{\rm phys}$ collects one-sided barriers of the same form as $\mathcal{B}$, each with its own weight, which enforce the thermodynamic stability, the positivity of the symmetry energy, the monotonicity of the pressure, the causality and the support of a $2M_\odot$ star. The four sets reported here satisfy all these requirements with a margin, so that every barrier is inactive and $\mathcal{P}_{\rm phys}=0$, as listed in Tab.~\ref{tab:loss}; parameter sets that violate one of them outright are rejected before the objective is evaluated.

The resulting contributions are listed in Tab.~\ref{tab:loss}. The M-R terms stay below $2.6$ in all the sets, the sum $\sum_j\chi^2_j$ running over two to four objects; the largest differences between the sets are carried by the HIC flow term, while the numerically dominant GW170817 term varies little.

\begin{table}[htbp]
\centering
\caption{Contributions to the total objective $\mathcal{L}_{\rm tot}$ of Eq.~\eqref{eq:Ltot} for the four parameter sets, in the same units. The values are those recorded at the end of the refinement and their sum reproduces the last column of Tab.~\ref{tab:sets}. They are not comparable across the sets, which are fitted to different M-R data.}
\label{tab:loss}
\begin{tabular}{lcccc}
\hline\hline
 & GQHD1 & GQHD2 & GQHD3 & GQHD4 \\
\hline
$\mathcal{P}_{\rm NM}$ & 0.56 & 1.30 & 1.68 & 2.02 \\
$\sum_j\chi^2_j$ & 1.58 & 2.54 & 0.55 & 1.84 \\
$\mathcal{P}_{\rm HIC}$ & 0.18 & 1.56 & 4.20 & 3.46 \\
$\mathcal{P}_{\rm reg}$ & 0.03 & 0.09 & 0.03 & 0.05 \\
$\mathcal{P}_{\rm phys}$ & 0 & 0 & 0 & 0 \\
$\mathcal{L}_{\rm GW}$ & 28.75 & 26.98 & 28.14 & 27.43 \\
\hline
$\mathcal{L}_{\rm tot}$ & 31.10 & 32.47 & 34.60 & 34.80 \\
\hline\hline
\end{tabular}
\end{table}


\section{Parameter sets}

Following Eq.~\eqref{eq:Ltot}, the locally optimized parameter sets for the four data scenarios of Tab.~\ref{tab:sets} were found and listed in Tab.~\ref{tab:Par}, and corresponding values of $\mathcal{L}_{\rm tot}$ have already been listed in Tabs.~\ref{tab:NM} and~\ref{tab:loss}. The reference models are listed in the same field convention; note that the sign of $g_\sigma$ is tied to the signs of the odd-power and mixed scalar couplings, so a set cannot be transferred to another RMF code by matching parameter names alone.
\begin{table}[htbp]
\centering
\caption{The estimated values of model parameters ($g_{\phi NN}$ denotes the coupling $g_\phi$ of the Lagrangian; rounded values are shown, the full-precision sets are available from the authors). $m_\sigma, b_5, b_9, g_2, g_6$ and $g_7$ are in unit of MeV. For the reference models the meson masses used in the calculation are $(m_\omega,m_\rho,m_{a_0},M_N)=(783,770,950,938)$~MeV for TM1 and $(782.5,763,980,939)$~MeV for NL3 and FSU-$\delta6.7$.}
\label{tab:Par}
\begin{tabular}{ccccccccc}
\hline\hline
Parameter & TM1 & NL3 & FSU-$\delta6.7$ & GQHD1 & GQHD2 & GQHD3 & GQHD4 & \\
\hline
$g_{\sigma NN}$ & 10.0 & $-10.2$ & $-10.2$ & $-10.7$ & 7.98 & 8.25 & 8.24 & \\
$g_{\omega NN}$ & 12.6 & 12.9 & 13.4 & 12.9 & 10.4 & 10.9 & 10.5 & \\
$g_{\rho NN}$ & 4.63 & 4.47 & 7.27 & 6.79 & 6.19 & 5.56 & 6.09 & \\
$g_{a_0 NN}$ & --- & --- & $-6.70$ & $-6.37$ & $-6.71$ & $-6.15$ & $-6.75$ & \\
$m_\sigma$ & 511 & 508 & 492 & 534 & 474 & 481 & 495 & \\
$b_5$ & --- & --- & --- & 120 & $-107$ & $-125$ & $-128$ & \\
$b_9$ & --- & --- & --- & 45.3 & $-1830$ & $-1800$ & $-1720$ & \\
$g_2$ & $-1430$ & 2060 & 1600 & 1680 & $-910$ & $-671$ & $-815$ & \\
$g_6$ & --- & --- & --- & 27.2 & $-1630$ & $-1960$ & $-1770$ & \\
$g_7$ & --- & --- & --- & 131 & $-53.7$ & $-46.4$ & $-27.7$ & \\
$b_3$ & --- & --- & --- & 65.7 & 237 & 252 & 280 & \\
$b_4$ & --- & --- & --- & 2.25 & 1.94 & 1.80 & 1.95 & \\
$b_6$ & --- & --- & 180 & 206 & $-26.0$ & $-5.23$ & $-19.6$ & \\
$b_7$ & --- & --- & --- & $-0.328$ & $-7.06$ & $-6.74$ & $-6.59$ & \\
$b_8$ & --- & --- & --- & 11.3 & $-333$ & $-316$ & $-313$ & \\
$g_3$ & 0.618 & $-28.9$ & 5.88 & 6.63 & 8.06 & 7.48 & 7.53 & \\
$g_4$ & --- & --- & --- & 1.93 & $-3.45$ & $-2.37$ & $-2.93$ & \\
$g_5$ & --- & --- & --- & 10.6 & 128 & 128 & 117 & \\
$c_3$ & 71.3 & --- & 172 & 166 & 11.4 & 9.56 & 1.22 & \\
$c_4$ & --- & --- & 204 & 194 & 0.0159 & 1.87 & 3.82 & \\
$d_3$ & --- & --- & --- & --- & 710 & 606 & 653 & \\
\hline\hline
\end{tabular}

\end{table}

\end{widetext}

\bibliography{GQHDRef}

\end{document}